\documentstyle[aps,epsfig,twocolumn]{revtex}

\title{Coupled quantum dots as quantum gates}

\author{
Guido Burkard$^{1*}$,
Daniel Loss$^{1\dagger}$,
and David P.\ DiVincenzo$^{2\ddagger}$}

\address{$^{1}$
Department of Physics and Astronomy,
University of Basel,\\ Klingelbergstrasse 82,
CH-4056 Basel, Switzerland}

\address{$^{2}$
IBM Research Division, T.J.\ Watson Research Center,\\
P.O.\ Box 218, Yorktown Heights, New York 10598}

\newcommand{\br}{{\bf r}}

\begin{document}

\twocolumn[\hsize\textwidth\columnwidth\hsize\csname @twocolumnfalse\endcsname

\maketitle

\begin{abstract}

We consider a new quantum gate mechanism based on electron spins in
coupled semiconductor quantum dots.  Such gates provide a general
source of spin entanglement and can be used for quantum computers.  We
determine the exchange coupling $J$ in the effective Heisenberg model
as a function of magnetic ($B$) and electric fields, and of the inter-dot
distance $a$ within the Heitler-London approximation of molecular
physics. This result is refined by using sp-hybridization, and by the
Hund-Mulliken molecular-orbit approach which leads to an extended
Hubbard description for the two-dot system that shows a remarkable
dependence on $B$ and $a$ due to the long-range Coulomb interaction.
We find that the exchange $J$
changes sign
at a finite field (leading to a pronounced jump in the magnetization)
and then decays exponentially.  The magnetization and the spin
susceptibilities of the coupled dots are calculated.  We
show that the dephasing due to nuclear spins in GaAs
can be strongly suppressed by dynamical nuclear spin polarization
and/or by magnetic fields.
\end{abstract}


\vskip2pc]
\narrowtext

\section{Introduction}

Semiconductor quantum dots, sometimes referred to as artificial atoms,
are small devices in which charge carriers are confined in all three
dimensions\cite{jacak}. The confinement is usually achieved by
electrical gating and/or etching techniques applied e.g. to a
two-dimensional electron gas (2DEG). Since the dimensions of quantum
dots are on the order of the Fermi wavelength,
their electronic spectrum consists of discrete energy levels which
have been studied in great detail in
conductance \cite{jacak,kouwenhoven} and
spectroscopy measurements\cite{jacak,ashoori,kotthaus}.
In GaAs heterostructures the number of electrons in the dots can be
changed one by one starting from zero\cite{tarucha}. Typical
laboratory magnetic fields ($B\approx 1\,{\rm T}$) correspond to
magnetic lengths on the order of $l_B\approx 10\,{\rm nm}$, being much
larger than the Bohr radius of real atoms
but of the same size as artificial atoms. As a consequence, the dot
spectrum depends strongly on the applied magnetic
field\cite{jacak,kouwenhoven,ashoori}.
In coupled quantum dots which can be considered to some extent as
artificial molecules, Coulomb blockade effects\cite{waugh} and
magnetization\cite{oosterkamp} have been observed as well as the
formation of a delocalized ``molecular state''\cite{blick}.

Motivated by the rapid down-scaling of integrated circuits, there has
been continued interest in classical logic devices made of
electrostatically coupled quantum dots\cite{nomoto}. More recently,
the discovery of new principles of computation based on quantum
mechanics\cite{deutsch85} has led to the idea of using coupled quantum
dots for quantum computation\cite{loss}; many other proposed
implementations have been explored, involving
NMR\cite{gershenfeld,cory97,jones}, trapped ions\cite{cirac}, cavity
QED\cite{turchette}, and Josephson junctions\cite{averin}.
Solid-state devices open up the possibility of fabricating large
integrated networks which would be required for realistic applications
of quantum computers.  A basic feature of the quantum-dot
scenario\cite{loss} is to consider
the {\it electron spin} ${\bf S}$ as the qubit (the qubit being the
basic unit of information in the quantum computer).  This stands in
contrast to alternative proposals also based on quantum
dots\cite{barenco95,landauer,hawrylak,rossi}, in which it is the {\it charge}
(orbital) degrees of freedom out of which a qubit is formed and
represented in terms of a pseudospin-1/2.  However, there are two
immediate advantages of real spin over pseudospin: First, the qubit
represented by a real spin-1/2 is always a well defined qubit; the
two-dimensional Hilbert space is the entire space available, thus
there are no extra dimensions into which the qubit state could ``leak"
\cite{footnote1}. Second, during a quantum computation phase coherence
of the qubits must be preserved. It is thus an essential advantage of
real spins that their dephasing times in GaAs can be on the order of
microseconds\cite{kikkawa}, whereas for charge degrees of freedom
dephasing times are typically much less, on the order of
nanoseconds\cite{huibers,footnotedephasing}.

In addition to a well defined qubit, we also need a controllable
``source of entanglement", i.e.  a mechanism by which two specified
qubits at a time can be entangled\cite{divincenzo95} so as to produce
the fundamental quantum XOR (or controlled-NOT) gate operation,
represented by a unitary operator $U_{\rm XOR}$\cite{g9}.  This can be
achieved by temporarily coupling two spins\cite{loss}.  As we will
show in detail below, due to the Coulomb interaction and the Pauli
exclusion principle the ground state of two coupled electrons is a
spin singlet, i.e. a highly entangled spin state.  This physical
picture translates into an exchange coupling $J(t)$ between the two
spins ${\bf S}_{1}$ and ${\bf S}_{2}$ described by a Heisenberg
Hamiltonian
\begin{equation}\label{Heisenberg}
H_{\rm s}(t)=J(t)\,\,{\bf S}_1\cdot{\bf S}_2.
\end{equation}
If the exchange coupling is pulsed such that $\int dtJ(t)/\hbar =
J_0\tau_s/\hbar = \pi$ (mod $2\pi$), the associated unitary time
evolution $U(t) = T\exp(i\int_0^t H_{\rm s}(\tau)d\tau/\hbar)$
corresponds to the ``swap'' operator $U_{\rm sw}$ which simply
exchanges the quantum states of qubit 1 and 2\cite{loss}. Furthermore,
the quantum XOR can be obtained\cite{loss} by applying the sequence
$\exp(i(\pi/2)S_1^z)\exp(-i(\pi/2)S_2^z)U_{\rm sw}^{1/2}\exp(i\pi
S_1^z)U_{\rm sw}^{1/2}\equiv U_{\rm XOR}$, i.e. a combination of
``square-root of swap'' $U_{\rm sw}^{1/2}$ and single-qubit rotations
$\exp(i\pi S_1^z)$, etc. Since $U_{\rm XOR}$ (combined with
single-qubit rotations) is proven to be a universal quantum
gate\cite{barenco95,divincenzo95}, it can therefore be used to
assemble any quantum algorithm.  Thus, the study of a quantum XOR gate
is essentially reduced to the study of the {\it exchange mechanism}
and how the exchange coupling $J(t)$ can be controlled experimentally.
We wish to emphasize that the switchable coupling mechanism described
in the following need not be confined to quantum dots: the same
principle can be applied to other systems, e.g. coupled atoms in a
Bravais lattice, overlapping shallow donors in semiconductors such
as P in Si\cite{kane}, and so on. The main reason to concentrate here
on quantum dots is that these systems are at the center of many
ongoing experimental investigations in mesoscopic physics, and thus
there seems to be reasonable hope that these systems can be made into
quantum gates functioning along the lines proposed here.

In view of this motivation we study in the following the spin dynamics
of two laterally coupled quantum dots containing a single electron
each. We show that the exchange coupling $J(B,E,a)$ can be controlled
by a magnetic field $B$ (leading to wave function compression), or by
an electric field $E$ (leading to level detuning), or by varying the
barrier height or equivalently the inter-dot distance $2a$ (leading to
a suppression of tunneling between the dots).  The dependence on these
parameters is of direct practical interest, since it opens the door to
tailoring the exchange $J(t)$ for the specific purpose of creating 
quantum gates.
We further calculate the static and dynamical magnetization responses
in the presence of perpendicular and parallel magnetic fields, and
show that they give experimentally accessible information about the
exchange $J$.  Our analysis is based on an adaptation of
Heitler-London and Hund-Mulliken variational techniques\cite{mattis} to
parabolically confined coupled quantum dots.  In particular, we
present an extension of the Hubbard approximation induced by the
long-range Coulomb interaction. We find a striking dependence of the
Hubbard parameters on the magnetic field and inter-dot distance which
is of relevance also for atomic-scale Hubbard physics in the presence
of long-range Coulomb interactions.  Finally, we discuss the effects of
dephasing induced by nuclear spins in GaAs and show that dephasing can
be strongly reduced by dynamically polarizing the nuclear spins and/or
by magnetic fields.

The paper is organized as follows. In Sec. II we introduce the model
for the quantum gate in terms of coupled dots. In Sec. III we calculate
the exchange coupling first in the Heitler-London and then
in the Hund-Mulliken approach. There we also discuss the
Hubbard limit and the new features arising from the long range nature of
the Coulomb interactions. In Sec. IV we consider the effects of
imperfections leading to dephasing and gate errors; in particular,
we consider dephasing resulting from nuclear spins in GaAs.
Implications for experiments on magnetization and spin
susceptibilities are presented in Sec. V, and Sec. VI contains
some concluding remarks on networks of gates with some suggestions
for single-qubit gates operated by local magnetic fields.
Finally, we mention that a preliminary account of some of the results
presented here has been given in Ref.\cite{divincenzo98a}.

\begin{figure}
\centerline{\psfig{file=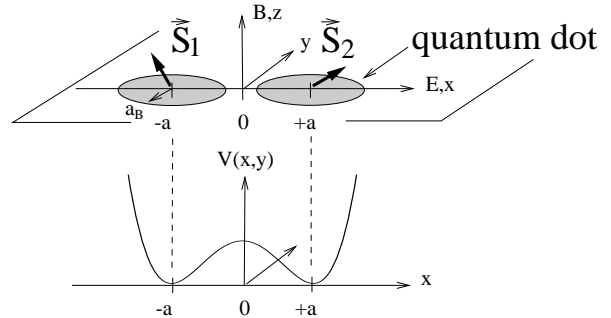,width=8cm}}\bigskip\bigskip
\caption{\label{system}
Two coupled quantum dots with one valence electron per dot.
Each electron is confined to the $xy$ plane.
The spins of the electrons in
dots $1$ and $2$ are denoted by ${\bf S}_1$ and ${\bf S}_2$. The
magnetic field $B$ is perpendicular to the plane, i.e. along the $z$
axis, and the electric field $E$ is in-plane and along the $x$
axis. The quartic potential is given in Eq.~(\ref{potential}) and is
used to model the coupling of two harmonic wells centered at $(\pm
a,0,0)$. The exchange coupling $J$ between the spins is a function of
$B$, $E$, and the inter-dot distance $2a$.}
\end{figure}

\section{Model for the quantum gate}

We consider a system of two laterally coupled quantum dots containing
one (conduction band) electron each, see Fig.~\ref{system}. It is
essential that the electrons are allowed to tunnel between the dots,
and that the total wave function of the coupled system must be
antisymmetric. It is this fact which introduces correlations
between the spins via the charge (orbital) degrees of freedom.
For definiteness we shall use in the following the parameter values
recently determined for single GaAs heterostructure quantum
dots\cite{tarucha} that are formed in a 2DEG; this choice is not
crucial for the following analysis but it allows us to illustrate our
analytical results with realistic numbers. The Hamiltonian for the
coupled system is then given by
\begin{eqnarray}
H &=& \sum_{i=1,2} h_i+C+H_{\rm Z} = H_{\rm orb} + H_{\rm Z},
\nonumber\\
h_i &=& \frac{1}{2m}\left({\bf p}_i-\frac{e}{c}{\bf A}(\br_i)
\right)^2+ex_iE+V(\br_i),\label{hamiltonian}\\
C&=&\frac{e^2}{\kappa\left| \br_1-\br_2\right|}.\nonumber
\end{eqnarray}
The single-particle Hamiltonian $h_i$ describes the electron dynamics
confined to the $xy$-plane. The electrons
have an effective mass $m$ ($m=0.067\, m_e$
in GaAs) and carry a spin-1/2 ${\bf S}_i$. 
The dielectric constant in GaAs is $\kappa = 13.1$. We allow for
a  magnetic field ${\bf B}= (0,0,B)$ applied along the $z$-axis
and which couples to the electron charge via the
vector potential ${\bf A}(\br) = \frac{B}{2}(-y,x,0)$.
We also allow for an electric  field $E$ applied in-plane along
the x-direction, i.e. along the line connecting the centers of the dots.
The coupling of the dots (which includes tunneling) is modeled by a quartic
potential,
\begin{equation}\label{potential}
V(x,y)=\frac{m\omega_0^2}{2}\left(\frac{1}{4 a^2}\left(x^2-a^2
\right)^2+y^2\right)\, ,
\end{equation}
which separates (for $x$ around $\pm a$) into two harmonic wells  of
frequency $\omega_0$, one for each dot, in the limit
of large inter-dot distance, i.e. for $2a\gg 2 a_{\rm B}$, where $a$ is
half the distance between the centers of the dots, and
$a_{\rm B}=\sqrt{\hbar/m\omega_0}$
is the effective Bohr radius of a single isolated harmonic well.
This
choice for the potential is motivated by the experimental
fact\cite{tarucha} that
the spectrum of single dots in GaAs is well described by a parabolic
confinement potential,
e.g. with $\hbar\omega_0 =3\,{\rm meV}$\cite{tarucha}.
We note that increasing (decreasing) the inter-dot distance
is physically equivalent to
raising (lowering) the inter-dot barrier, which can be
achieved experimentally by e.g.
applying a gate voltage between the dots\cite{waugh}.
Thus, the effect of such gate voltages is described in our model
simply by a change of the inter-dot distance $2a$.
We also note that it is only for simplicity that we choose the two dots
to be exactly identical, no qualitative changes will occur in the
following analysis if the dots are only approximately equal and
approximately of
parabolic shape.

The (bare) Coulomb interaction between the two electrons is
described by $C$. The screening length $\lambda$ in almost depleted regions
like few-electron quantum dots can be expected to be much larger than the
bulk 2DEG screening length (which is about $40\,{\rm nm}$ in GaAs).
Therefore, $\lambda$ is large compared to the size of the coupled system,
$\lambda\gg 2a\approx 40\,{\rm nm}$
for small dots, and we will consider the limit of unscreened
Coulomb interaction
($\lambda/a\gg 1$) throughout this work.

The magnetic field $B$ also couples to the electron spins via the
Zeeman term $H_{\rm Z}=g\mu_{\rm B} \sum_i{\bf B}_i\cdot{\bf S}_i$,
where $g$ is the effective g-factor ($g\approx -0.44$ for GaAs), and
$\mu_{\rm B}$ the Bohr magneton.  The ratio between the Zeeman splitting and
the relevant orbital energies is small for all $B$-values of interest
here; indeed, $g\mu_{\rm B} B/\hbar\omega_0\lesssim 0.03$, for $B\ll
B_0=(\hbar\omega_0/\mu_{\rm B})(m/m_e)\approx 3.5\,{\rm T}$,
and $g\mu_{\rm B} B/\hbar\omega_{\rm L}\lesssim 0.03$, for $B\gg B_0$,
where $\omega_{\rm L}=eB/2mc$ is the Larmor
frequency, and where we used $\hbar\omega_0 =3\,{\rm meV}$. Thus, we
can safely ignore the Zeeman splitting when we discuss the orbital
degrees of freedom and include it later into the effective spin
Hamiltonian.  Also, in the few-electron system we are dealing with,
spin-orbit effects can be completely neglected since $H_{\rm
so}/\hbar\omega_0\approx 10^{-7}$, where $H_{\rm
so}=(\omega_0^2/2mc^2){\bf L}\cdot{\bf S}$ is the spin-orbit coupling
of an electron in a parabolic confinement
potential\cite{divincenzo98a}.  This has the important implication
that dephasing effects induced e.g. by potential or charge
fluctuations in the surroundings of the isolated dots can couple only
to the charge of the electron so that they have very small influence
on the phase coherence of the isolated spin itself (for dephasing
induced by coupling the dots see Sec. \ref{dephasingsection}).  It is
for this reason that it is preferable to consider dots containing
electrons instead of holes, since holes will typically have a sizable
spin-orbit interaction\cite{jacak}.

Finally, we assume a low-temperature description where $kT\ll
\hbar\omega_0$, so that we can restrict ourselves to the two lowest
orbital eigenstates of $H_{\rm orb}$, one of which is symmetric
(spin singlet) and the other one antisymmetric (spin triplet).
In this reduced (four-dimensional) Hilbert space, $H_{\rm orb}$
can be replaced by the effective Heisenberg spin Hamiltonian
Eq.~(\ref{Heisenberg}), $H_{\rm s}=J{\bf S}_1\cdot{\bf S}_2$, 
where the exchange energy $J=\epsilon_{\rm t}-\epsilon_{\rm s}$
is the difference between the triplet and
singlet energy which we wish to calculate. The above model cannot be
solved in an analytically closed form. However, the analogy between atoms
and quantum dots (artificial atoms) provides us with a powerful set of
variational methods from molecular physics for finding $\epsilon_{\rm t}$
and $\epsilon_{\rm s}$. Note that the typical energy scale $\hbar\omega_0\approx
{\rm meV}$ in our quantum dot is about a thousand times smaller than
the energies (${\rm Ry}\approx {\rm eV}$) in a hydrogen atom, whereas
the quantum dot is larger by about the same factor. This is important
because their size makes quantum dots much more susceptible to
magnetic fields than atoms.  In analogy to atomic physics, we call the
size of the electron orbitals in a quantum dot the Bohr radius,
although it is determined by the confining potential rather than by
the Coulomb attraction to a positively charged nucleus. For harmonic
confinement $a_{\rm B}=\sqrt{\hbar/m\omega_0}$ is
about $20\,{\rm nm}$ for $\hbar\omega_0 =3\,{\rm meV}$.

\section{Exchange energy}

\subsection{Heitler-London approach}

\begin{figure}
\centerline{\psfig{file=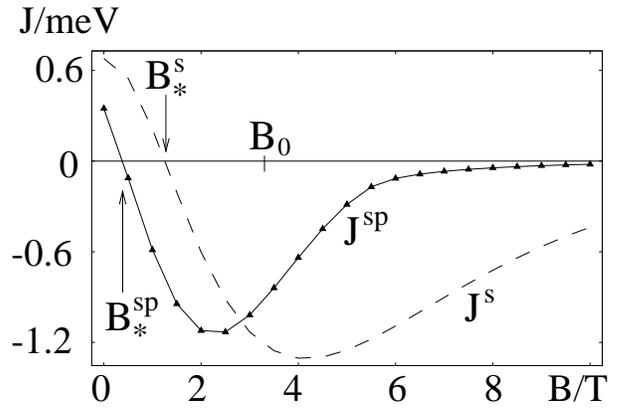,width=8cm}}\bigskip\bigskip
\caption{\label{exchange}Exchange energy $J$ in units of
meV plotted against the magnetic field $B$ (in units of Tesla), as
obtained from the s-wave Heitler-London approximation (dashed line),
Eq.~(\ref{J}), and the  result from the improved
sp-hybridized Heitler-London approximation (triangles) which is obtained
numerically as explained in the text.
Note that the qualitative behavior
of the two curves is similar, i.e. they both
have zeroes, the s-wave approximation at $B_*^{\rm s}$,
and the sp-hybridized approximation at $B_*^{\rm sp}$, and also both curves
vanish exponentially for large fields.
$B_0=(\hbar\omega_0/\mu_{\rm B})(m/m_e)$ denotes the crossover
field to magnetically dominated confining ($B\gg B_0$).
The curves are given for a confinement energy $\hbar\omega_0=3\,{\rm meV}$
(implying for the Coulomb parameter $c=2.42$), and
inter-dot distance $a=0.7\, a_{\rm B}$.}
\end{figure}

\begin{figure}
\centerline{\psfig{file=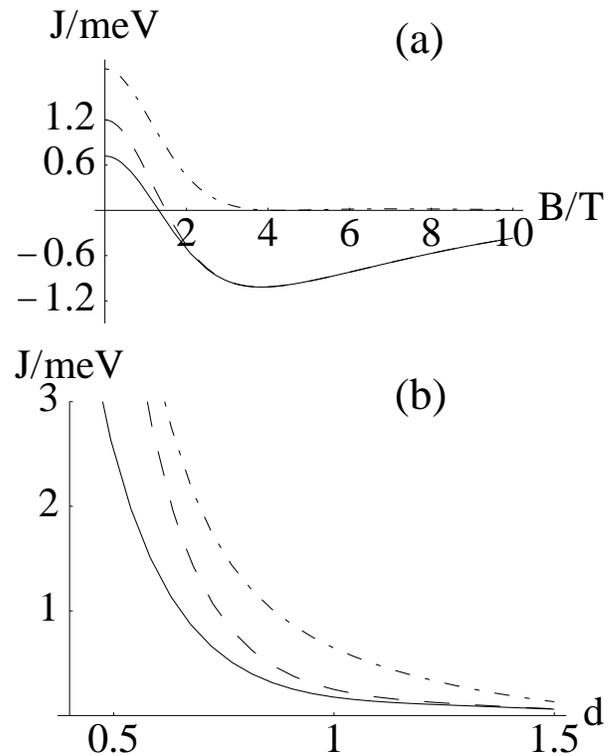,width=8cm}}\bigskip\bigskip
\caption{\label{plots}The exchange coupling $J$ obtained from
Hund-Mulliken (full line), Eq.~(\ref{HMresult}), and from the extended
Hubbard approximation (dashed line), Eq.~(\ref{Hubbard}). For
comparison, we also plot the usual Hubbard approximation where the
long-range interaction term $V$ is omitted, i.e.  $J=4t_{\rm
H}^2/U_{\rm H}$ (dashed-dotted line).  In (a), $J$ is plotted as a
function of the magnetic field $B$ at fixed inter-dot distance
($d=a/a_{\rm B}=0.7$), and for $c=2.42$, in (b) as a function of
inter-dot distance $d=a/a_{\rm B}$ at zero field ($B=0$), and again
$c=2.42$.  For these parameter values, the s-wave Heitler-London $J$,
Eq.~(\ref{J}), and the Hund-Mulliken $J$ (full line) are almost
identical.}
\end{figure}

We consider first the Heitler-London approximation, and then refine this
approach by including hybridization as well as double occupancy
in a Hund-Mulliken approach, which will finally lead us to an extension of
the Hubbard description. We will see, however,  that
the qualitative features of $J$ as a  function of the control parameters
are already
captured by the simplest Heitler-London approximation for the artificial
hydrogen molecule described by Eq.~\ref{hamiltonian}.
In this approximation, one starts from
single-dot ground-state orbital wavefunctions $\varphi(\br)$ and combines
them into the (anti-) symmetric  two-particle orbital state vector
\begin{equation}
|\Psi_{\pm}\rangle
= \frac{|12\rangle
\pm |21\rangle}{\sqrt{2(1\pm S^2)}},
\end{equation}
the positive (negative) sign
corresponding to the spin singlet (triplet) state,
and $S=\int d^2r\varphi_{+a}^{*}(\br)\varphi_{-a}(\br)=\langle 2|1\rangle$
denoting the overlap of the right and left orbitals.
A non-vanishing overlap implies that the electrons tunnel
between the dots (see also Sec. \ref{HM}).
Here, $\varphi_{ -a}(\br)=\langle \br| 1\rangle$ and $\varphi_{+a}(\br)
=\langle \br|2\rangle$ denote the one-particle orbitals
centered at $\br=(\mp a,0)$, and $|ij\rangle = |i\rangle |j\rangle$ are
two-particle product states. The exchange energy is then obtained
through $J = \epsilon_{\rm t}-\epsilon_{\rm s} = 
\langle\Psi_{-}|H_{\rm orb}|\Psi_{-}\rangle -
\langle\Psi_{+}|H_{\rm orb}|\Psi_{+}\rangle$.
The single-dot orbitals for harmonic
confinement in two dimensions in a perpendicular magnetic field are the
Fock-Darwin states \cite{Fock}, which are the usual harmonic oscillator
states, magnetically compressed by a factor $b=\omega/\omega_0 =
\sqrt{1+\omega_{\rm L}^2/\omega_0^2}$, where $\omega_{\rm L} = eB/2mc$
denotes the Larmor frequency. The ground state
(energy $\hbar\omega=b\hbar\omega_0$) centered at the origin is
\begin{equation}
\varphi(x,y) = \sqrt{\frac{m \omega}{\pi\hbar}} e^{-m\omega
\left(x^2+y^2\right)/2\hbar}.
\end{equation}
Shifting the single particle orbitals to $(\pm a,0)$ in the
presence of a magnetic field we obtain $\varphi_{\pm a}(x,y) = \exp(\pm
iya/2l_B^2)\varphi(x\mp a,y)$. The phase factor involving the magnetic
length $l_B=\sqrt{\hbar c/eB}$ is due to the gauge transformation ${\bf
A}_{\pm a}=B(-y,x\mp a,0)/2\rightarrow {\bf A}=B(-y,x,0)/2$. The matrix
elements of $H_{\rm orb}$ needed to calculate $J$ are found by adding and
subtracting the harmonic potential centered at $x=-(+)\, a$ for electron
$1(2)$ in $H_{\rm orb}$, which then takes the form $H_{\rm orb}=h^0_{-a}({\bf
r}_1)+h^0_{+a}({\bf r}_2)+W+C$, where $h^0_{\pm a}({\bf r}_i)=({\bf
p}_i-e{\bf A}({\bf r}_i)/c)^2/2m+m\omega^2((x_i\mp a)^2+y_i^2)/2$ is the
Fock-Darwin Hamiltonian shifted to $(\pm a,0)$, and
$W(x,y)=V(x,y)-m\omega^2((x_1+a)^2+(x_2-a)^2)/2$. We obtain
\begin{equation}\label{Jformal}
J = \frac{2S^2}{1-S^4}\left(\langle 12|C+W|12\rangle-\frac{{\rm Re}
\langle 12|C+W|21\rangle}{S^2}\right),
\end{equation}
where the overlap becomes $S=\exp(-m\omega a^2/\hbar-a^2\hbar/4l_B^4m\omega)$.
Evaluation of the matrix elements of $C$ and $W$ yields
(see also \cite{divincenzo98a})
\begin{eqnarray}
J&=&\frac{\hbar\omega_0}{\sinh\left(2d^2(2b-1/b)\right)}\Bigg[c\sqrt{b}
\Bigg(e^{-bd^2}{\rm I_0}(bd^2)\nonumber\\
&-& e^{d^2 (b-1/b)}{\rm I_0}(d^2
(b-1/b))\Bigg)+\frac{3}{4b}\left(1+bd^2\right)\Bigg],\label{J}
\end{eqnarray}
where we introduce the dimensionless distance $d=a/a_{\rm B}$, and ${\rm
I_0}$ is the zeroth order Bessel function.  The first and second terms
in Eq.~(\ref{J}) are due to the Coulomb interaction $C$, where the
exchange term enters with a minus sign.  The parameter
$c=\sqrt{\pi/2}(e^2/\kappa a_{\rm B})/\hbar\omega_0$ ($\approx 2.4$, for
$\hbar\omega_0=3\,$meV) is the ratio between Coulomb and
confining energy. The last term comes from the confinement potential
$W$.  The result $J(B)$ is plotted in Fig.~\ref{exchange} (dashed
line).  Note that typically $|J/\hbar\omega_0|\lesssim 0.2$. Also, we
see that $J>0$ for $B=0$, which must be the case for a two-particle
system that is time-reversal invariant\cite{mattis}.
The most remarkable feature of $J(B)$, however, is the change of sign
from positive to negative at $B=B_*^{\rm s}$, which occurs over a wide
range of parameters $c$ and $a$. This singlet-triplet crossing occurs
at about $B_*^{\rm s}=1.3\,{\rm T}$ for $\hbar\omega_0=3\,{\rm meV}$
 ($c=2.42$) and $d=0.7$. The transition from antiferromagnetic ($J>0$) to
ferromagnetic ($J<0$) spin-spin coupling with increasing magnetic
field is caused by the long-range Coulomb interaction, in particular
by the negative exchange term, the second term in Eq.~(\ref{J}).  As
$B\gg B_0$ ($\approx 3.5\,{\rm T}$ for $\hbar\omega_0=3\,$meV),
the magnetic field compresses the orbits by a factor $b\approx
B/B_0\gg 1$ and thereby reduces the overlap of the wavefunctions,
$S^2\approx\exp(-2 d^2(2b-1/b))$, exponentially strongly. Similarly, the
overlap decays exponentially for large inter-dot distances, $d\gg 1$.
Note however, that this exponential suppression is partly compensated
by the exponentially growing exchange term $\langle 12|C|21\rangle/S^2\propto
\exp(2d^2(b-1/b))$. As a result, the exchange coupling $J$ decays
exponentially as $\exp(-2d^2b)$ for large $b$ or $d$, as shown in
Fig.~\ref{plots}b for $B=0$ ($b=1$).
Thus, the exchange coupling $J$ can be tuned
through zero and then suppressed to zero by a magnetic field in a
very efficient way.
We  note that our Heitler-London approximation breaks down
explicitly (i.e. $J$ becomes negative even when $B=0$) for certain
inter-dot distances when $c$ exceeds $2.8$.
Finally, a similar singlet-triplet crossing as function of the magnetic
field has been found in {\it single} dots with two
electrons\cite{wagner}.

The exchange energy $J$ also depends on the applied electric field
$E$.  The additional term $e(x_1+x_2)E$ in the potential merely shifts
the one-particle orbitals by $\Delta x = eE/m\omega_0^2$, raising the
energy of both the singlet and triplet states.  Since the singlet
energy turns out to be less affected by this shift than the triplet,
the exchange energy $J$ increases with increasing $E$,
\begin{equation}
J(B,E) = J(B,0)+ \frac{\hbar\omega_0}{\sinh(2d^2(2b-1/b))}\frac{3}{2}
\frac{1}{d^2}\left(\frac{eEa}{\hbar\omega_0}\right)^2,
\label{efield}
\end{equation}
the increase being proportional to $m\omega_0^2(\Delta x)^2$. 
[We note that this increase of $J(B,E)$ is qualitatively 
consistent with what one finds from  a standard two-level approximation of a 1D
double-well potential (with $J(B,0)$ being the effective tunnel splitting) in the
presence of a bias given by $eEa$.] The variational Ansatz leading to
Eq.~(\ref{efield}) is expected to remain accurate as long as $J(B,E) -
J(B,0)\lesssim J(B,0)$; for larger
$E$-fields the levels of the dots get completely detuned and the
overlap of the wavefunctions (i.e. the coherent tunneling) between the
dots is suppressed. Of course, a sufficiently large electric field
will eventually force both electrons on to the same dot, which is the
case when $eEa$ exceeds the on-site repulsion $U (\gg J(B,E=0)$, see
below).  However, this situation, which would correspond to a quantum
dot helium\cite{pfannkuche}, is not of interest in the present
context.  Conversely, in case of dots of different size (or shape) where the
energy levels need not be aligned a priori, an appropriate electric field can be
used to match the levels of the two dots, thus allowing coherent
tunneling even in those systems.  Recent conductance
measurements\cite{blick} on coupled dots of different size 
(containing several electrons) with electrostatic tuning
have revealed clear evidence for a delocalized molecular state.

A shortcoming of the simple approximation described above is that
solely ground-state single-particle orbitals were taken into account
and mixing with excited one-particle states due to interaction is
neglected. This approximation is self-consistent if $J\ll
\Delta\epsilon$, where $\Delta\epsilon$ denotes the single-particle
level separation between the ground state and the first excited
state. We find $|J/\Delta\epsilon| < 0.25$ at low fields $B\le
1.75\,{\rm T}$, therefore $J(B)$ is at least qualitatively correct in
this regime. At higher fields $|J/\Delta\epsilon|\approx 1$,
indicating substantial mixing with higher orbitals. An improved
Heitler-London variational Ansatz is obtained by introducing
sp-hybridized single-dot orbitals (in analogy to molecular physics),
i.e.~$\phi = \varphi_{\rm s} + \alpha\varphi_{{\rm
p}x}+i\beta\varphi_{{\rm p}y}$, where $\varphi_{\rm s}=\varphi$ is the
s-orbital introduced above, $\varphi_{{\rm p}q}=\sqrt{2\over \pi}m\omega
q\exp(-m\omega r^2/2\hbar)/\hbar$, $q=x,y$, are the lowest two Fock-Darwin
excited states (at zero field) with angular momentum $|\ell|=1$, and
$\alpha$ and $\beta$ are real variational parameters to be determined
by minimization of the singlet and triplet energies
$\epsilon_{\rm s,t}(\alpha,\beta)$, which is done numerically. The
$\varphi_{{\rm p}q}$ are chosen to be real, they are however not
eigenstates of the single-particle Hamiltonian, which are
$\varphi_{{\rm p}x}\pm i\varphi_{{\rm p}y}$ (with eigenenergy
$2\hbar\omega\pm
\hbar\omega_{\rm L}$). Note that while $\epsilon_{\rm s,t}$ decrease
only by $\approx 1\%$ due to hybridization, 
the relative variation of $J=\epsilon_{\rm t}-\epsilon_{\rm s}$
can still be substantial.  Nevertheless, the
resulting exchange energy $J^{\rm sp}$ (Fig.~\ref{exchange}) is only
quantitatively different from the pure s-wave result $J\equiv J^{\rm
s}$, Eq.~(\ref{J}). At low fields, $J^{\rm sp}<J^{\rm s}$ and the
change of sign occurs already at about $B_*^{\rm sp}\simeq 0.4\,{\rm
T}<B_*^{\rm s}$. At high fields, $J^{\rm sp}$ shows a much more
pronounced decay as a function of $B$.

Being a completely orbital effect, the exchange interaction
between spins of course competes with the Zeeman coupling $H_Z$ of the
spins to the magnetic field. In our case, however, the Zeeman energy $H_Z$
is small and exceeds the exchange energy (polarizing the spins) only in a
narrow window (about $0.1\,{\rm T}$ wide) around $B_*^{\rm sp}$ and again
for high fields ($B>4\,{\rm T}$).

\subsection{Hund-Mulliken approach and Hubbard Limit}
\label{HM}

We turn now to the Hund-Mulliken method of molecular orbits \cite{mattis}
which extends the Heitler-London approach by including also the two
doubly occupied states, which both are spin singlets. This extends
the orbital Hilbert space from two to four dimensions.
First, the single particle states have to be orthonormalized,
leading to the states $\Phi _{\pm a}=(\varphi_{\pm a}-g\varphi_{\mp
a})/\sqrt{1-2Sg+g^2}$, where $S$ again denotes the overlap of
$\varphi_{-a}$ with $\varphi_{+a}$ and $g=(1-\sqrt{1-S^2})/S$. Then,
diagonalization of
\begin{equation}\label{matrix}
H_{\rm orb} = 2\epsilon + \left(\begin{array}{cccc}
U&X&-\sqrt{2}t_{\rm H}&0\\
X&U&-\sqrt{2}t_{\rm H}&0\\
-\sqrt{2}t_{\rm H}&-\sqrt{2}t_{\rm H}&V_+&0\\
0&0&0&V_-
\end{array}\right)
\end{equation}
in the space spanned
by $\Psi^{\rm d}_{\pm a}(\br_1,\br_2)=\Phi_{\pm a}(\br_1)\Phi_{\pm
a}(\br_2)$, $\Psi^{\rm
s}_{\pm}(\br_1,\br_2)=[\Phi_{+a}(\br_1)\Phi_{-a}(\br_2)\pm
\Phi_{-a}(\br_1)\Phi_{+a}(\br_2)]/\sqrt{2}$ yields the eigenvalues
$\epsilon_{{\rm s}\pm} = 2\epsilon + U_{\rm H}/2+V_+\pm\sqrt{U_{\rm
H}^2/4+4t_{\rm H}^2}$, $\epsilon_{{\rm s} 0} = 2\epsilon + U_{\rm H} - 2X
+ V_+$ (singlet), and $\epsilon_{\rm t} = 2\epsilon + V_-$ (triplet),
where the quantities
\begin{eqnarray}
\epsilon &=& \langle\Phi_{\pm a}|h^0_{\pm a}|\Phi_{\pm a}\rangle , \nonumber\\
t_{\rm H} &=& t- w = \langle\Phi_{\pm
a}|h^0_{\pm}|\Phi_{\mp a}\rangle -
\langle\Psi^{\rm s}_+|C|
\Psi^{\rm d}_{\pm a}\rangle/\sqrt{2}, \nonumber\\
V &=& V_- - V_+ = \langle\Psi^{\rm
s}_-|C|\Psi^{\rm s}_-\rangle-
\langle\Psi^{\rm s}_+|C|\Psi^{\rm s}_+\rangle, \label{matrixelements} \\ 
X &=& \langle \Psi^{\rm d}_{\pm a}|C|\Psi^{\rm d}_{\mp a}\rangle, \nonumber\\
U_{\rm H} &=& U-V_+ +X \nonumber\\
&=& \langle\Psi^{\rm d}_{\pm a}
|C|\Psi^{\rm d}_{\pm a}\rangle-\langle\Psi^{\rm s}_+|C|\Psi^{\rm
s}_+\rangle+\langle\Psi^{\rm d}_{\pm a}|C|\Psi^{\rm d}_{\mp a}\rangle,\nonumber
\end{eqnarray}
all depend on the magnetic field $B$. The exchange energy is the gap
between the lowest singlet and the triplet state
\begin{equation}
J = \epsilon_{\rm t}-\epsilon_{{\rm s}-}=V - \frac{U_{\rm H}}{2} +
\frac{1}{2}\sqrt{U_{\rm H}^2 + 16t_{\rm H}^2}.
\label{HMresult}
\end{equation}
In the standard Hubbard approach for short-range Coulomb interactions (and
without
$B$-field) \cite{mattis} $J$ reduces to $-U/2 + \sqrt{U^2+ 16t^2}/2$,
where $t$ denotes the hopping matrix element,
and $U$ the on-site repulsion (cf. Eq. (\ref{matrixelements})).
Thus, $t_{\rm H}$ and $U_{\rm H}$ are the extended hopping matrix element
and the on-site repulsion, resp., renormalized by  long-range Coulomb
interactions. The remaining two singlet energies $\epsilon_{{\rm s}+}$ and
$\epsilon_{{\rm s}0}$
are separated from $\epsilon_{\rm t}$ and $\epsilon_{{\rm s}-}$ by a gap of
order $U_{\rm H}$ and are therefore neglected for the study of
low-energy properties.
The evaluation of the matrix elements is straightforward but lengthy,
and we give the results in Appendix \ref{appendix}.
Typically, the ``Hubbard ratio'' $t_{\rm H}/U_{\rm H}$ is less than
$1$, e.g., if $d=0.7$, $\hbar\omega_0=3\,$meV, and $B=0$,
we obtain $t_{\rm H}/U_{\rm H}=0.34$, and it decreases with 
increasing $B$. Therefore, we are in an extended Hubbard
limit, where $J$ takes the form
\begin{equation}
J = \frac{4t_{\rm H}^2}{U_{\rm H}}+V.
\label{Hubbard}
\end{equation}
The first term has the form of the standard Hubbard 
approximation \cite{fradkin} (invoked previously\cite{loss})
but with
$t_{\rm H}$
and $U_{\rm H}$ being renormalized by long-range Coulomb interactions.
The second term $V$ is new and
accounts for the difference in Coulomb energy between the
singly occupied singlet and triplet states $\Psi^{\rm s}_{\pm}$. It is
precisely this $V$
that makes $J$ negative for high magnetic fields, whereas $t_{\rm
H}^2/U_{\rm H}>0$ for all values of $B$ (see Fig.~\ref{plots}a).
Thus, the usual Hubbard approximation (i.e. without $V$) would not give
reliable
results, neither for the $B$-dependence (Fig.~\ref{plots}a)
nor for the dependence on
the inter-dot distance $a$ (Fig.~\ref{plots}b) \cite{footnote:atoms}.
Since only the singlet space has
been enlarged, it is clear that we obtain a lower singlet energy
$\epsilon_{\rm s}$ than that from the s-wave Heitler-London calculation, 
but the same
triplet energy $\epsilon_{\rm t}$, and therefore 
$J=\epsilon_{\rm t}-\epsilon_{\rm s}$
exceeds the s-wave Heitler-London result, Eq. (\ref{J}).
However, the on-site Coulomb
repulsion
$U\propto c$ strongly suppresses the doubly occupied states $\Psi^{\rm
d}_{\pm a}$ and already for the value of $c=2.4$ (corresponding
to $\hbar \omega_0=3 $meV) we
obtain almost perfect agreement with the s-wave Heitler-London result
(Fig.~\ref{exchange}).
For large fields, i.e. $B\gg B_0$,
the suppression becomes even stronger ($U\propto \sqrt{B}$) because
the electron orbits become compressed with increasing $B$ and two electrons on
the same dot are confined to a smaller area leading to an increased Coulomb
energy.

\section{Dephasing and Quantum gate errors}
\label{dephasingsection}

We allow now for imperfections and discuss first the dephasing
resulting from coupling to the environment, and then address briefly
the issue of errors during the quantum-gate operation. We have
already pointed out that dephasing in the charge sector will have
little effect on the (uncoupled) spins due to the smallness of the
spin-orbit interaction. Similarly, the dipolar interaction between the
qubit spin and the surrounding spins is also minute, it can be
estimated as $(g\mu_{\rm B})^2/a_{\rm B}^3\approx 10^{-9}$ meV.
Although both couplings are extremely small they will eventually lead
to dephasing for sufficiently long times. We have described such
weak-coupling dephasing in terms of a reduced master equation
elsewhere\cite{loss}, and we refer the interested reader to this work.
Since this type of dephasing is small it can be eliminated by error
correction schemes\cite{preskill}.

Next, we consider the dephasing due to nuclear spins in GaAs
semiconductors, where both Ga and As possess a nuclear spin
$I=3/2$. There is a sizable hyperfine coupling between the
electron-spin ($s=1/2$) and all the nuclear spins in the quantum dot
which might easily lead to a flip of the electron spin and thus cause
an error in the quantum computation. We shall now estimate this effect
and show that it can be substantially reduced by spin polarization or
by a field.  We consider an electron spin ${\bf S}$ in contact with
$N$ nuclear spins ${\bf I}^{(i)}$ in the presence of a magnetic field
$B\parallel z$. The corresponding Hamiltonian is given by
$H=A{\bf
S}\cdot{\bf I}+b_z S_z+\tilde{b}_z I_z=H_0+V$ ,
where
\begin{equation}
H_0=A S_z I_z +b_z S_z+\tilde{b}_z I_z ,\,\, V=A(S_+ I_- + S_- I_+)/2.
\end{equation}
Here, $A$ is a hyperfine coupling, ${\bf I}=\sum_{i=1}^N {\bf I}^{(i)}$ is the
total nuclear spin, and $b_z=g \mu_{\rm B} B_z$, $\tilde{b}_z=g_N \mu_N B_z$
($g_N$ and $\mu_N$ denote the nuclear g factor and magneton). Consider
the initial eigenstate $|i\rangle$ of $H_0$, which we will
consider to be one basis vector for the qubit, where the electron spin
is up (in the $S_z$ basis), and the nuclear spins are in a product
state of $I_z^{(i)}$-eigenstates with total $I_z=pNI$ ($-1\le p \le
1$), i.e.\ in a state with polarization $p$ along the $z$-axis; here,
$p=\pm 1$ means that the nuclear spins are fully polarized in positive
(negative) $z$-direction, and $p=0$ means no polarization. Due to the
hyperfine coupling the electron spin can flip (i.e. dephase) with the
entire system going into a
final state $|k\rangle$ which is again a product state but now with
the electron-spin
down, and, due to conservation of total spin,
the $z$-component $I_z^{(k)}$ of one and only one nuclear spin having
increased by
$2s=1$.  All final states $|k\rangle$ are degenerate and again
eigenstates of
$H_0$ with eigenenergy $E_f$. We will consider this process now within
time-dependent
perturbation theory  and up to second order in $V$.
The energy difference between
initial and final states amounts to
$E_i-E_f\approx 2s[A(pIN+s)+b_z]$, where
we use that $b_z\gg \tilde{b}_z$.  For the reversed process with
an electron-spin flip from down to up but with the same initial
polarization for the nuclear spins the energy difference is $\approx
-2s[A(pIN-s)+b_z]$.
The total transition probability to leave
the initial state $|i\rangle$
after time $t$ has elapsed is then
\begin{equation}
P_{i}(t)=\left(\frac{2\sin((E_f-E_i)t/2\hbar)}{E_f-E_i}\right)^2
\sum_{k(\neq i)}|\langle k|V|i \rangle|^2 .
\end{equation}
We interpret this total transition probability $P_{i}(t)$ as the
degree of decoherence caused by spin-flip processes over time $t$.
Now, $|\langle k|V|i\rangle |^2= A^2[I(I+1)-I_z^{(k)}(I_z^{(k)}+1)]/4$.
Assuming some  distribution of the nuclear spins
we can replace this matrix element by its average value (denoted by
brackets) where
$\sqrt{\langle(I_z^{(k)})^2\rangle}$ describes then the variance of
the mean value $\langle{I_z^{(k)}}\rangle=pI$. E.g. a Poissonian
distribution gives $|\langle k|V|i\rangle |^2\approx
A^2[I(I+1)-pI(pI+1)]/4$, in which case the matrix element
vanishes for full polarization parallel to the electron-spin (i.e.
$p=1$), as required by conservation of total spin.
$P_{i}(t)$ is strongly suppressed for final states for which
$t_0\equiv 2\pi\hbar/|E_i-E_f|\ll t$, which simply reflects
conservation of energy.
In particular, for a substantial nuclear polarization,
i.e. $p^2 N\gg 1$, $P_{i}(t)$ oscillates in time but with the
vanishingly small amplitude $1/p^2N$ (for $B=0$).
We can estimate $N$ to be on the order of the number of atoms per quantum
dot, which is about $10^5$.
Such a situation with $p^2N\gg 1$ can be established by dynamically
spin-polarizing the nuclear spins (Overhauser effect) e.g. via
optical pumping\cite{dobers} or
via spin-polarized currents at the edge of a 2DEG\cite{dixon}.
This gives rise to an effective nuclear field
$B_{\rm n}=ApNI/g\mu_{\rm B}$ which is reported to be as
large as  $B_{\rm n}^* =4\,{\rm T}$ in
GaAs  (corresponding to $p=0.85$)\cite{dixon} and which has a lifetime on
the order of minutes\cite{dobers}.
Alternatively, for unpolarized nuclei with $p=0$ but a field $B$
in the Tesla range,
the amplitude of $P_{i}(t)$ vanishes as $(AIN/g\mu_{\rm B} B)^2/N
\approx(B_{\rm n}^*/B)^2/N\ll 1$.
For  $B$ or $B_{\rm n}=1\,{\rm T}$ the oscillation
frequency $1/t_0$ of $P_{i}(t)$ is about $10$ GHz.
Thus,  spin flip processes and hence dephasing due to nuclear
spins can be strongly
suppressed, either by dynamically
polarizing the nuclear spins and/or by applying a magnetic field $B$.
The remaining dephasing effects (described again by a
weak-coupling master equation\cite{loss}) should then be small enough to be
eliminated by error correction.

We now address the imperfections of the quantum gate operation.  For
this we note first that, for the purpose of quantum computing, the
qubits must be coupled only for the short time of switching $\tau_s$,
while most of the time there is to be no coupling between the dots.  We
estimate now how small we can choose $\tau_s$. For this we consider a
scenario where $J$ (initially zero) is adiabatically switched on and
off again during the time $\tau_s$, e.g. by an electrical gate
by which we lower and then
raise again the barrier $V(t)$ between the dots (alternatively, we
can vary $B$, $a$, or $E$).
A typical frequency scale during switching is given by
the exchange energy (which results in the coherent tunneling between the
dots) averaged over
the time interval of switching,
$\overline{J}=(1/\tau_s)\int_0^{\tau_s} dt\, J(t)$. Adiabaticity then
requires that
many coherent oscillations (characterized approximately by $\overline{J}$)
have to take place in
the double-well system while the control parameter $v$ = $V$, $B$, $a$,
or $E$ is being changed, i.e.\
$1/\tau_s\approx |\dot{v}/v|\ll {\overline J}/\hbar$.
If this criterion is met, we can use our
equilibrium analysis to calculate $J(v)$ and then simply
replace $J(v)$ by $J(v(t))$ in case of a time-dependent control
parameter $v(t)$ \cite{footnotecriterion}.
Note that this is compatible
with the requirement needed for the XOR operation, $J\tau_s/\hbar=n\pi$, n odd,
if we choose $n\gg 1$.  Our method of calculating $J$ is
self-consistent if $J\ll\Delta \epsilon$, where $\Delta \epsilon$
denotes the single-particle level spacing. The combination of both
inequalities yields $1/\tau_s\ll {\overline J}/\hbar\ll \Delta \epsilon/\hbar$,
i.e.\ no
higher-lying levels can be excited during the switching.  Finally,
since typically $J\approx 0.2\,$ meV we see that $\tau_s$ should not be
smaller than about $50\,$ps.
Now, during the time $\tau_s$ spin and
charge couple and thus dephasing in the charge sector described by
$\tau_\phi^c$ can induce dephasing of spin via an uncontrolled
fluctuation $\delta J$ of the exchange coupling.
However, this effect is again
small, it can be estimated to be on the order of $\tau_s/\tau_\phi^c
\sim 10^{-2}$, since even for large dots $\tau_\phi^c$ is reported to
be on the order of nanoseconds\cite{huibers}. This seems to be a rather
conservative estimate and one can expect the spin dephasing to
be considerably smaller since not every charge dephasing
event will affect the spin.
Finally, weak dephasing of the effective spin Hamiltonian during switching
has been described elsewhere\cite{loss} in terms of a weak-coupling master
equation which accounts explicitly for decoherence of the spins during the
switching process. Based on this analysis\cite{loss}, the probability for
a gate error per gate operation (described by ${\cal K}_2$ in Eq.~(13)
of \onlinecite{loss}) is estimated to be approximately $\tau_s/\tau_\phi^c
\sim 10^{-2}$ or better (see above).

\begin{figure}
\centerline{\psfig{file=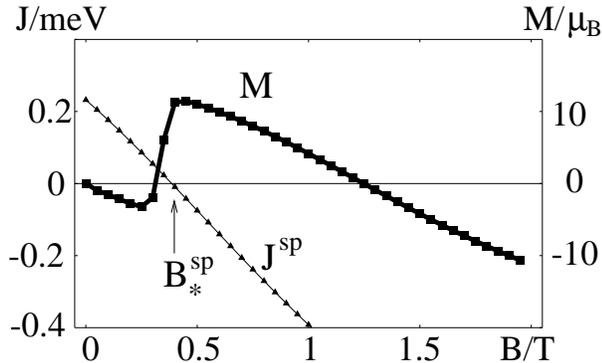,width=8cm}}\bigskip\bigskip
\caption{\label{magnetization}The equilibrium magnetization $M$
(box-shaped symbols) in units of Bohr magnetons $\mu_{\rm B}$ as a function
of magnetic field. $M$ is obtained numerically from the sp-hybridized
Heitler-London approximation. Note that the magnetization exhibits a
jump at the field value $B_*^{\rm sp}$ for which the exchange $J^{\rm
sp}$ (triangle symbols) changes sign.  At the left and right hand side
of the jump the negative slope of $M(B)$ indicates orbital
diamagnetism. The temperature for this plot is $T=0.2\,$K, while as
before $\hbar\omega_0=3\,{\rm meV}$ and $a=0.7\, a_{\rm B}$.}
\end{figure}

\section{Experimental Implications}

Coherent coupling between the states of neighboring dots is the
keystone of our proposal for quantum gate operation, and experimental
probes of this coupling will be very interesting to explore.  The
effect of the dot-dot coupling manifests itself in the level structure,
which could be measured non-invasively with spectroscopic
methods\cite{ashoori,kotthaus}. An alternative way is to measure the
static magnetization in response to a magnetic field $B$ which is
applied along the $z$-axis.  This equilibrium magnetization is given
by $M=g\mu_{\rm B}{\rm Tr}(S^z_1+S^z_2) e^{-(H_s +H_Z)/kT}$, where $H_s$ is given
in Eq.~(\ref{Heisenberg}), and $H_{\rm Z}=g\mu_{\rm B} \sum_i{\bf
B}_i\cdot{\bf S}_i$ is the Zeeman term.  It is straightforward to
evaluate $M$, and in Fig.~\ref{magnetization} we plot $M$ as a
function of $B$ for a typical temperature $T=0.2\,$K.
The exchange $J^{\rm sp}(B)$ is also shown in Fig.~\ref{magnetization}.
Both $J^{\rm sp}(B)$ and $M$ are the results of the sp-hybridized
Heitler-London approximation. We note that the equilibrium
magnetization $M(B)$ is strongly dominated by the orbital response
(via the exchange $J$); we find a diamagnetic response (negative slope
of $M$) for $B<B_*^{\rm sp}$ which is followed by a pronounced jump in
the magnetization at the field $B_*^{\rm sp}$ followed again by a
diamagnetic response. Experimental observation of this jump would give
evidence for the existence of the predicted singlet-triplet
level-crossing at $B_*^{\rm sp}$, and such measurements would allow
one to ``map out" $J$ around the point where it can be tuned to zero,
e.g. by also varying the barrier between the dots. The magnetic moment
produced by the orbital motion of the electrons in one pair of coupled
quantum dots at the peak ($B=B_*^{\rm sp}$) is around $10 \mu_{\rm B}$
(see Fig.~\ref{magnetization}). This signal could be further amplified
by using an ensemble of pairs of coupled quantum dots.

A further way to get experimental information about the exchange
coupling would be to measure the spin response to an ac magnetic field
(in the linear-response regime), described by the dynamical spin
susceptibilities $\chi^{pq}_{mn}(\omega ) =
(i/\hbar)\int_0^{\infty}dt\exp(i\omega t)\langle [
S^p_m(t),S^q_n(0)]\rangle$, where $m,n=1,2$, and $p,q=x,y,z$.  Being
interested in the spin response only, we assume this ac field to be
applied in plane so that there is no orbital response (for a sufficiently
weak field with no subband mixing). We see then that all the transverse
spin susceptibilities $\chi^{p\neq q, q}_{mn}$ vanish, and we are left
with the longitudinal ones only, where $\chi^{xx}_{mn}=\chi^{yy}_{mn}=
\chi^{zz}_{mn}\equiv \chi_{mn}$ due to
the rotational symmetry of $H_{\rm s}$.  It is sufficient to consider the
dissipative part, $\chi_{mn}^{''}(\omega )={\rm Im} \chi_{mn}(\omega)$,
for which we obtain $\chi_{11}^{''}=\chi_{22}^{''}=-\chi_{12}^{''}=
-\chi_{21}^{''}=-(\pi/4)f(J , B) [\delta(\hbar\omega+J)-
\delta(\hbar\omega-J)]$, where $f(J, B)=(e^{ J/kT}-1)/(1+{e^{ J/kT}}+
2\ \cosh (g\mu_{\rm B} B/kT))$.  Also, due to conservation of total spin,
the total response, $\chi_{1j}+\chi_{2j}$, as well as the response 
to a spatially uniform field, $\chi_{i1}+\chi_{i2}$, vanish. 
Thus, to observe the spin susceptibilities calculated here one needs
to apply the fields locally or to measure the spin of a dot separately;
both cases could be realized e.g. by atomic or magnetic force
microscopes (see also below, where we briefly discuss local fields 
produced by field gradients).

\section{Concluding remarks}

We end with a few comments on a network of coupled quantum dots in the
presence of fields (see also Ref.\onlinecite{loss}). In a set-up with
only one quantum gate (i.e. two quantum dots) the gate operation can be
performed using uniform magnetic fields (besides electric gates),
while in a quantum computer with many gates, which have to be
controlled individually, local magnetic fields are indispensable,
especially for the single-qubit gates\cite{loss,footnote3}.
However, we emphasize that it is not necessary that every single quantum
dot in a network is directly addressable with a local magnetic field.
Indeed, using ``swap'' operations $U_{\rm sw}$, any qubit-state can be
transported to a region where the single-qubit gate operation is
performed, and then back to its original
location, without disturbing this or other qubits.
In one possible mode of operation a constant field $B_*$, defined by
$J(B_*)=0$, is applied, while smaller time-dependent {\it local}
fields then control the gate operations.  We can envision local fields
being achieved by a large number of techniques: with neighboring
magnetic dots\cite{loss}, closure domains, a grid of current-carrying
wires below the dots, tips of magnetic or atomic force microscopes, or
by bringing the qubit into contact (by shifting the dot via electrical
gating) with a region containing magnetic moments or nuclear spins
with different hyperfine coupling (e.g. AlGaAs instead of GaAs)--and
others.  A related possibility would be to use magnetic field
gradients.  Single-qubit switching times of the order of
$\tau_s\approx 20\,$ps require a field of $1\,$T, and for an inter-dot
distance $2a\approx 30\,$nm, we would need gradients of about $1\,{\rm
T}/30\,{\rm nm}$, which could be produced
with commercial disk reading/writing heads.  [The operation of
several XOR gates via magnetic fields also requires gradients of
similar magnitude.]
Alternatively, one could use an ac magnetic field
$B_{\rm ac}$ and apply electron spin resonance (ESR) techniques to
rotate spins with a single-qubit switching time (at resonance)
$\tau_s\approx \pi \hbar/B_{\rm ac}$.
To address the dots of an array individually with ESR, a magnetic field
gradient is needed which can be estimated as follows.  Assuming a
relative ESR linewidth of $1\%$ and again $2a=30\,$nm we find about
$B_{\rm ac}\cdot 10^4\,{\rm cm}^{-1}$.  Field gradients in excitation
sequences for NMR up to $2\cdot 10^4\,{\rm G}/{\rm cm}$ have been
generated\cite{cory98} which allows for $B_{\rm ac}\approx 1\,$G.
The resulting switching times, however, are rather long, on the order of
$100$ ns, and larger field gradients would be desirable.
Finally, such ESR techniques could be employed to obtain information
about the effective exchange values $J$: the exchange coupling between
the spins leads to a shift in the spin resonance frequency which we
found to be of the order of $J/\hbar$ by numerical analysis\cite{unpublished}.

To conclude, we have calculated the exchange energy $J(B,E,a)$ between spins of
coupled quantum dots (containing one electron each) as a function of
magnetic and electric fields and inter-dot distance using the
Heitler-London, hybridized Heitler-London, and Hund-Mulliken
variational approach.  We have shown that $J(B,E,a)$ changes
sign (reflecting a singlet-triplet crossing) with increasing $B$ field
before it vanishes exponentially.  Besides being of fundamental
interest, this dependence opens up the possibility to use coupled
quantum dots as quantum gate devices which can be operated by magnetic
fields and/or electric gates (between the dots) to produce
entanglement of qubits.

\acknowledgments
We would like to thank J. Kyriakidis, S. Shtrikman,
and E. Sukhorukov for useful discussions.  This work has been
supported in part by the Swiss National Science Foundation.

\appendix
\section{Hund-Mulliken matrix elements}\label{appendix}

Here, we list the explicit expressions for the matrix elements
defined in Eqs.~(\ref{matrix}) and (\ref{matrixelements}) as 
a function of the dimensionless
inter-dot distance $d=a/a_{\rm B}$ and the magnetic compression
factor $b=\sqrt{1+\omega_{\rm L}^2/\omega_0^2}$ where
$\omega_{\rm L}=eB/2mc$. The single-particle matrix elements
are given by
\begin{eqnarray}
\epsilon &=& \frac{3}{32}\frac{1}{b^2 d^2}
           + \frac{3}{8} \frac{S^2}{1 - S^2}
             \left(\frac{1}{b} + d^2\right) + b, \\
t &=&  \frac{3}{8} \frac{S}{1 - S^2} \left(\frac{1}{b} + d^2\right),
\end{eqnarray}
where we used $S=\exp(-d^2(2b - 1/b))$.
The (two-particle) Coulomb matrix elements can be expressed as
\begin{eqnarray}
V_+ &=& N^4 \left( 4 g^2 (1 + S^2) F_1 + (1 + g^2)^2 F_2 \right.\nonumber\\
                      &+& \left. 4 g^2 F_3 -  16 g^2 F_4\right),\\
V_- &=& N^4 (1-g^2)^2 (F_2  -  S^2 F_3), \\
U   &=& N^4 \left( (1 + g^4 + 2 g^2 S^2)F_1 + 2 g^2 F_2\right.\nonumber\\
                   &+& \left.2 g^2 S^2 F_3 - 8 g^2 F_4 \right),\\
X   &=& N^4 \left[\left( (1 + g^4) S^2 + 2 g^2 \right) F_1 + 2 g^2 F_2 \right.\nonumber\\
                      &+& \left.2 g^2 S^2 F_3 - 8 g^2 F_4 \right],\\
w   &=& N^4 \left( - g (1 + g^2) (1 + S^2) F_1 - g (1 + g^2) F_2 \right.\nonumber\\
               &-& \left. g (1 + g^2) S^2 F_3 + (1 + 6 g^2 + g^4) S F_4 \right),
\end{eqnarray}
with $N=1/\sqrt{1-2Sg+g^2}$ and $g=(1-\sqrt{1-S^2})/S$.
Here, we make use of the functions
\begin{eqnarray}
F_1 &=& c \sqrt{b}, \\
F_2 &=& c \sqrt{b} \, e^{-b d^2} {\rm I}_0\left(b d^2\right),\\
F_3 &=& c \sqrt{b} \, e^{d^2 (b - 1/b)} {\rm I}_0 \left(d^2(b - 1/b)\right),\\
F_4 &=& c \sqrt{b} \, e^{-d^2/4b} \times\nonumber\\
    &\times&\sum_{k=-\infty}^{\infty}(-1)^k
         {\rm I}_{2k}\left(\frac{d^2}{4}(2b-1/b)\right)
         {\rm I}_{2k}\left(i\frac{d^2}{2}\sqrt{b^2 - 1}\right),\nonumber\\
\end{eqnarray}
where ${\rm I}_n$ denotes the Bessel function of $n$-th order. For
our purposes, we can neglect terms with $|k|>1$ in the sum in $F_4$,
since for $\hbar\omega_0=3\,{\rm meV}$, $B<30\,{\rm T}$, and $d=0.7$ 
the relative error introduced by doing so is less than 1{\%}.


\begin{thebibliography}{99}
\bibitem{footnote1} Such leakage can happen e.g.  in the switching
process of single qubits or by coupling two qubits together etc.  and
can easily lead to uncontrollable errors.  This concern is especially
relevant in quantum dots where the energy level spacing is (nearly)
uniform (in contrast to real atoms) so that the levels defining the
qubit are of similar scale as the separation to neighboring energy
levels.
\bibitem{footnotedephasing} The dephasing times of
Refs.\onlinecite{kikkawa,huibers} are both measured in GaAs
semiconductors which involve many electrons. It would be highly
desirable to get direct experimental information about dephasing times
in isolated quantum dots of low filling as considered here.
\bibitem{footnote:atoms} We note that the significant changes due to
Coulomb long-range interactions  are valid down to the 
scale of real atoms. Since atomic orbitals and the harmonic orbitals
used here behave similarly (for $B=0$), we expect to find qualitatively similar
results for real molecules (as found here for coupled dots)
especially as regards the
effect of Coulomb long-range interactions on $t_{\rm H},U_{\rm H}, J$
and their dependence on the inter-atomic distance $a$.
\bibitem{footnotecriterion}
If during the change of $v(t)$ the total spin remains conserved, no
transitions between the instantaneous singlet and triplet eigenstates can be
induced during the switching. Thus, the singlet and triplet states evolve
independently of each other, and the condition on adiabatic switching involves
$\Delta \epsilon$ (instead of $J$), i.e. we only need to require that
$1/\tau_s\approx |\dot v/v|\ll 
\overline{\Delta \epsilon}/\hbar$, which would be less restrictive.
Also, only $\int_0^{\tau_s}dt J(t)$
and not $J(t)$ itself is needed for the gate operation. 
Therefore, the
adiabaticity criterion given in the text, while being sufficient, need not be
really necessary. 
However, the complete
analysis of the time-dependent problem in terms of variational wave functions
is beyond the scope of the present work and will be addressed elsewhere.
\end{thebibliography}
\end{document}